\documentclass[twocolumn,english]{revtex4-1}
\usepackage[T1]{fontenc}
\usepackage[latin9]{inputenc}
\usepackage{amsmath}
\usepackage{amssymb}
\usepackage{graphicx}

\makeatletter
\relax

\makeatother

\usepackage{babel}
\begin{document}
\title{
\global\long\def\bra{\langle}%
\global\long\def\ket{\rangle}%
\global\long\def\ddt{\mathrm{\frac{d}{dt}}}%
\global\long\def\cm{\mathrm{cm^{-1}}}%
}
\title{Modeling Non-Reversible Molecular Internal Conversion Using the Time-dependent
Variational Approach with $\text{sD}_{2}$ \textit{Ansatz}}
\author{Mantas Jaku\v{c}ionis\textsuperscript{1}, Tomas Mancal\textsuperscript{2},
Darius Abramavi\v{c}ius\textsuperscript{1}}
\affiliation{\textsuperscript{1}Institute of Chemical Physics, Vilnius University,
Sauletekio Ave. 9-III, LT-10222 Vilnius, Lithuania}
\affiliation{\textsuperscript{2}Faculty of Mathematics and Physics, Charles University,
Ke Karlovu 5, 121 16 Prague, Czech Republic}
\begin{abstract}
Effects of non-linear coupling between the system and the bath vibrational
modes on the system internal conversion dynamics are investigated
using the Dirac-Frenkel variational approach with the defined $\text{sD}_{2}$
\textit{ansatz}. It explicitly accounts for the entangled system electron-vibrational
wavepacket states, while the bath quantum harmonic oscillator (QHO)
states are expanded in a superposition of coherent states (CS). Using
a non-adiabatically coupled three-level model, we show that quadratic
system-bath coupling induces non-reversible internal conversion when
the bath QHO wavepacket representation is highly non-Gaussian. The
quadratic coupling results in a broadened and asymmetrically squeezed
bath QHO wavepackets in the coordinate-momentum phase space. Additionally,
we found that computational effort can be reduced using degenerate
CSs to represent the initial bath wavepackets.
\end{abstract}
\maketitle

\section{Introduction\label{sec:Introduction}}

Function of many biological molecular systems is tightly connected
to the process of energy relaxation in their electronic or vibrational
(or both) manifolds. The problem of unraveling photo-excitation energy
relaxation pathways is relevant on a wide range of molecular spatial
scales: from the smallest molecular aggregates, consisting of just
a couple of molecules \citep{Balevicius2016,BaleviciusJr2019a,Meneghin2018,Staleva-Musto2019},
to photosynthetic complexes involving tens or hundreds of pigments
\citep{Thyrhaug2018,Fox2017,Maly2016}. Generally, due to a high number
of degrees of freedom (DOF) involved, brute-force numerical simulations
of even the smallest systems are challenging. The standard approach
to overcome this challenge is to apply the reduced (density operator)
description within the theory of open quantum systems \citep{Valkunasa,Weiss2012,Breuer2002a}.
In this description the most relevant electronic and vibrational DOFs
of the problem constitute the observable \textsl{system}, while all
the rest of DOFs are treated as a part of the fluctuating thermal
reservoir, the \textsl{bath}. When such a distinction is associated
with a small parameter characterizing the interaction strength between
the system and the bath, relatively simple perturbative approaches
are sufficient to describe energy relaxation phenomena. In a more
general case, division into the system and its bath is only formal,
as electronic states may be strongly coupled to both the vibrational
states of the system and those of the bath. As a result, excitation
energy exchange mechanisms between different states have to be modeled
non-perturbatively.

Dynamics of open quantum systems can be obtained by a broad range
techniques. In recent years, formally exact, but relatively expensive,
approach of the Hierarchical equations of motion \citep{Tanimura1989,Tanimura1990,Kreisbeck2012,Balevicius2013}
has gained popularity. Among the perturbative methods, various forms
of the Redfield theory \citep{Redfield1957b,Redfield1965} of the
weak system-bath coupling, and the Förster type of methods \citep{Forster1948,May2011a,Dinh2016,Seibt2017}
for the weak resonance coupling limit, still play an essential role
in understanding biologically relevant energy transfer and relaxation
processes. Among the phenomenological approaches, the Lindblad equations
\citep{Lindblad1976b,Breuer2002a} with their convenient formal properties
provide basis for cheap and reliable modelling. All the above mentioned
techniques are based on the density operator description, however,
for the same purposes one can also directly treat the wavefunction
itself, \textit{i.e.}, to expand electronic and vibrational states
of the model in a chosen electron-vibrational state basis. One family
of formally exact wavefunction approaches are based on the multi-configuration
time-dependent Hartree method (MCTDH) \citep{Meyer1990,Beck2000}
and include its multi-layer \citep{Wang2003,Wang2008}, Gaussian-based
\citep{Ronto2013,Richings2015} extensions. Additionally, methods
of coupled coherent states \citep{Shalashilin2000,Shalashilin2004},
its generalization to non-adiabatic dynamics -- multiconfigurational
Ehrenfest \citep{Shalashilin2010,Makhov2017}, variational multiconfigurational
Gaussians \citep{Worth2003,Worth2008}, iterative real-time path integral
\citep{Weiss2008a,Thorwart2009} are also available.

Wavefunction technique utilizing the time-dependent Dirac-Frenkel
variational principle with a trial wavefunction (\textit{ansatz})
based on the Davydov $\text{D}_{2}$ \textit{ansatz} for the molecular
chain soliton theory \citep{Davydov1979,Scott1991} is also being
developed. It models dynamics of both the system and the bath vibrational
DOFs approximatelly by representing vibrational states using coherent
states (CSs). Accuracy of the technique have been shown to improve
by considering more general variants of the Davydov $\text{D}_{2}$
\textit{ansatz}, \textit{i.e.}, $\text{D}_{1}$ \textit{ansatz} \citep{Somoza2017a}
or by using intermediatelly complex $\text{D}_{1.5}$ \textit{ansatz}
\citep{Werther2018a}. Still, the greatest improvement came by considering
a trial wavefunction made of a linear superposition of Davydov $\text{D}_{2}$
\textit{ansatz} ($\text{multi-D}_{2}$) and its more complex $\text{multi-D}_{1}$
variant \citep{Zhou2015a,Zhou2016,Wang2016}. Simulations of exciton
and polaron dynamics and non-linear optical spectra of molecular aggregates
\citep{Chorosajev2014a,Huynh2013}, light harvesting complexes \citep{Chen2015a},
also, dynamics of a simplified pyrazene excitation relaxation through
conical intersection \citep{Chen2019}, have proven the technique
to be a potent and flexible tool for simulating open quantum system
excitation energy relaxation dynamics and both the linear and non-linear
spectra.

In the present work, we extend this approach by considering non-linear
system-bath coupling terms to allow for vibrational energy exchange
between the system and the bath vibrational DOFs within the normal
mode description, using the modified $\text{multi-D}_{2}$ \textit{ansatz}
within Born-Oppenheimer approximation (BOA). Entangled system electron-vibrational
states were included formally exactly using coordinate representation,
while the bath vibrational states were represented by a superposition
of coherent states. We show that non-reversible internal conversion
requires highly non-Gaussian bath wavepacket representation, for the
system vibrational mode energy dissipation to bath to occur, and that
non-linear system-bath coupling results in a broadened and asymmetrically
squeezed bath QHO wavepackets along its coordinate and momentum axes.

The rest of the paper is structured as follows: In Section (\ref{sec:Theory})
we specify a general interacting system-bath model and give a brief
overview of the Dirac-Frenkel variational principle, while the derivation
of model equations of motion are presented in Supplementary Information.
In Section (\ref{sec:Results_Disc}) we present dynamics of a simulated
excitation relaxation between non-adiabatically coupled three-level
model with anharmonic potential energy surfaces (PES) and non-linear
vibrational-bath coupling. We also investigate effects of the initially
degenerate CSs representation on the dynamics convergence and discuss
the relevance of our approach. Conclusions are provided in Section
(\ref{sec:Conclusion}).

\section{Theory\label{sec:Theory}}

We consider a simple quantum system consisting of electronic and vibrational
DOFs. Electronic states and specific \textit{internal} vibrational
DOFs constitute the observable system (a molecule). This system is
coupled to a fluctuating bath, composed of a large number of \emph{external}
vibrational DOFs of molecule environment, \textit{e.g.}, vibrations
of polymeric matrix, proteins, solvent molecules, etc. Here, and throughout
the paper, for the internal and external vibrational manifold we will
use dimensionless coordinates $x$, $\chi$ and momenta $p$, $\rho$,
respectivelly, and also set the reduced Planck constant equal to one
($\hbar=1$).

The Hamiltonian operator $\hat{H}$ of the described quantum system
can be written as a sum of the system operator $\hat{H}_{\text{S}}$,
the bath operator $\hat{H}_{\text{B}}$, electronic-bath interaction
operator $\hat{H}_{\text{E-B}}$ and internal vibrational-bath interaction
operator $\hat{H}_{\text{V-B}}$
\begin{align}
\hat{H} & =\hat{H}_{\text{S}}+\hat{H}_{\text{B}}+\hat{H}_{\text{E-B}}+\hat{H}_{\text{V-B}}\ .\label{eq:ham-oper}
\end{align}
The system consists of $N$ electronic states $|n\rangle\left(n=0,1,\ldots,N\right)$,
with $\varepsilon_{n}$ representing the ground-excited state transition
$\left(|0\rangle\rightarrow|n\rangle\right)$ energies. Each electronic
state $|n\rangle$ is attached to $Q$ internal vibrational modes.
Vibrational modes $q=1,2,\ldots,Q$ are characterized by the generalized
$Q$-dimensional PES $V\left(\boldsymbol{x}\right)$, where $\boldsymbol{x}=\left(x_{1},x_{2},\ldots,x_{Q}\right)$
is a $Q$-dimensional space point. PESs attached to different electronic
states may differ, thus, the surface associated with the state $|n\rangle$
will be labeled as a diagonal term $V_{nn}\left(\boldsymbol{x}\right)$.

To represent quantum states of a vibrational mode $q$, we use the
coordinate representation for which the action of coordinate operator
$\hat{x}_{q}$ on coordinate state $|x_{q}\rangle$ has the eigenvalue
$x_{q}$: $\hat{x}_{q}|x_{q}\rangle=x_{q}|x_{q}\rangle$. For each
mode $q$ we consider coordinate states with eigenvalues from the
interval $x_{q}\in\left[x_{q}^{\text{min}},x_{q}^{\text{max}}\right]$
with equidistant spacing\textbf{ $\delta x_{q}$} between the states.
States $|x_{q}\rangle$ form a Q-dimensional space states $|\boldsymbol{x}\rangle\equiv|x_{1}\rangle|x_{2}\rangle\ldots|x_{Q}\rangle$
with orthonormality condition $\langle x_{q}|x_{q'}\rangle=\delta_{q,q'}\delta\left(x_{q}-x_{q^{'}}\right)$,
where $\delta_{a,b}$ and $\delta\left(c\right)$ are Kronecker and
Dirac delta functions, respectively. We will refer to the generalized
system electronic-vibrational states
\begin{align}
|n,\boldsymbol{x}\rangle & \equiv|n\rangle|x_{1}\rangle|x_{2}\rangle\ldots|x_{Q}\rangle\ ,
\end{align}
as \textit{vibronic} states.

It is well established that the PESs $V_{nn}\left(\boldsymbol{x}\right)$
of different molecular electronic states can get close to each other
in their energies (the avoided crossing region) or even cross each
other (the conical intersection) \citep{Domcke2004}, allowing for
non-radiative excitation relaxation between different electronic states.
Such a transition is called the internal conversion. During the internal
conversion, the molecule traverses to the lower energy electronic
state $|n\rangle\rightarrow|m<n\rangle$ with the excess energy $\varepsilon_{n}-\varepsilon_{m}>0$
being converted into the molecular vibrational energy (reverse transition
is also possible). Such a process is facilitated by the non-adiabatic
interaction between PESs of electronic state $|n\rangle$ and $|m\rangle$,
\textit{i. e.}, by $Q$-dimensional off-diagonal PES term $V_{nm}\left(\boldsymbol{x}\right)=V_{mn}\left(\boldsymbol{x}\right)$.

The complete Hamiltonian of the system therefore is defined as
\begin{align}
\hat{H}_{\text{S}} & =\sum_{n}\varepsilon_{n}|n\rangle\langle n|+\sum_{q}\frac{\omega_{q}}{2}\hat{p}_{q}^{2}+\sum_{n,m}\hat{V}_{nm}\left(\boldsymbol{x}\right)|n\rangle\langle m|\ ,\label{eq:H_sys}
\end{align}
where $\hat{p}_{q}=-\text{i}\frac{\partial}{\partial x_{q}}$ is the
system-related momentum operator. The Hamiltonian operator of the
bath is simply that of QHOs
\begin{equation}
\hat{H}_{\text{B}}=\sum_{p}\frac{w_{p}}{2}\left(\hat{\rho}_{p}^{2}+\hat{\chi}_{p}^{2}\right)\ .
\end{equation}

We assume that before an external excitation, the system is in its
electronic ground state $|0\rangle$ and both the system and bath
vibrational DOFs are in a state of thermodynamic equilibrium. All
system-bath coupling terms will be defined with respect to this pre-excitation
equilibrium state, therefore, system in its ground state is stationary
(\textit{i.e.} it is effectively not influenced by the bath DOFs in
any way).

System-bath interactions will be modelled via two mechanisms. First,
the excited electronic states energies are to be modulated by the
bath fluctuations. This will be modeled using the shifted PES model
\citep{May2011a}, \textit{i.e.}, surfaces $V_{nn}\left(\boldsymbol{x}\right)$
are shifted along the bath oscillator reaction coordinates $\chi_{p}$
by $s_{np}$, relative to the minimum of the $V_{00}\left(\boldsymbol{x}\right)$.
For convenience, we choose displacements to be directed in the positive
$\chi_{p}$ axis. In the regime of linear electronic-bath interaction,
the coupling is described by the Hamiltonian operator
\begin{equation}
\hat{H}_{\text{E-B}}=\sum_{n,p}w_{p}\left(\frac{1}{2}s_{np}^{2}-s_{np}\hat{\chi}_{p}\right)|n\rangle\langle n|\ ,\label{eq:E-B cpl}
\end{equation}
where the first term represents a shift of the electronic state $|n\rangle$
excitation energy, while the second term induces dynamical electronic
state $|n\rangle$ energy modulation via fluctuating bath coordinate
$\chi_{p}$. This additional excitation energy shift of $|n\rangle$
state is usually termed the bath reorganization energy $\Lambda_{n}^{\text{bath}}$,
and it is often merged with the $\varepsilon_{n}$. We keep them separate
in this work.

Second, to allow vibrational energy relaxation in the system (vibrational
energy exchange between the system and the bath), we include interaction
terms between vibrational system coordinates and the bath modes up
to a second order. Then the corresponding vibrational-bath interaction
Hamiltonian reads as
\begin{equation}
\hat{H}_{\text{V-B}}=\sum_{n,q,p}\left(k_{nqp}^{\left(1,1\right)}\hat{x}_{q}\hat{\chi}_{p}+k_{nqp}^{\left(1,2\right)}\hat{x}_{q}\hat{\chi}_{p}^{2}+k_{nqp}^{\left(2,1\right)}\hat{x}_{q}^{2}\hat{\chi}_{p}\right)|n\rangle\langle n|\ ,\label{eq:vibr-bath_cpl}
\end{equation}
where matrices $k_{nqp}^{\left(1,1\right)}$, $k_{nqp}^{\left(1,2\right)}$,
$k_{nqp}^{\left(2,1\right)}$ define interaction strengths between
vibrational mode $q$ and $p$ when system is in electronic state
$|n\rangle$ for different order coupling terms, indicated by the
supercript.

Statistical properties of the bath can be defined for a single specific
system-bath coupling term. For example, according to Eq. (\ref{eq:E-B cpl}),
the excited electronic state $|n\rangle$ energy modulation by the
bath fluctuations can be characterized by the spectral density function
$C_{n}''\left(w\right)$ \citep{Valkunasa,Weiss2012,Breuer2002a},
which can be defined in terms of $V_{nn}\left(\boldsymbol{x}\right)$
displacements $s_{np}$ as
\begin{align}
C_{n}''\left(v\right) & =\frac{\pi}{2}\sum_{p}s_{np}^{2}w_{p}^{2}\left(\delta\left(v-w_{p}\right)-\delta\left(v+w_{p}\right)\right).\label{eq:def-spd}
\end{align}
 Here $v$ is the parameter of the spectral density function $-\infty<v<\infty$,
while $w_{p}>0$. Notice, that this form leads to $C_{n}''\left(v\right)=-C_{n}''\left(-v\right)$.
The corresponding bath reorganization energy is then given by
\begin{align}
\Lambda_{n}^{\text{bath}} & =\int_{0}^{\infty}\frac{\text{d}v}{\pi}\frac{C_{n}''\left(v\right)}{v}\equiv\frac{1}{2}\sum_{p}s_{np}^{2}w_{p}.\label{eq:def-reorg}
\end{align}
 The constant $\pi$ comes from normalization of the Fourier transform.
Combining Eq. (\ref{eq:def-spd}) and (\ref{eq:def-reorg}), the bath
oscillator displacement absolute values $\left|s_{np}\right|$ can
be expressed as 
\begin{align}
\left|s_{np}\right| & =\frac{1}{w_{p}}\sqrt{\frac{2C_{n}''\left(w_{p}\right)\text{d}v}{\pi}},
\end{align}
where $\text{d}v$ is the discretization step size.

To define other system-bath coupling matrices, we further  assume
for simplicity that both electronic and vibrational DOFs of the system
interact with the same DOFs of the bath (the same external vibrational
modes), the interaction strength matrix $k_{nqp}^{\left(\beta\right)}$
elements, with $\beta=\left\{ 1,1\right\} ,\left\{ 1,2\right\} ,\left\{ 2,1\right\} $,
will then be expressed in terms of displacements $s_{np}$ (see Ref.
\citep{Jakucionis2018})
\begin{equation}
k_{nqp}^{\left(\beta\right)}=\gamma^{\left(\beta\right)}\frac{w_{p}\left|s_{np}\right|}{\sqrt{2}}\ ,
\end{equation}
 for all $q$, where $\gamma^{\left(\beta\right)}$ is a dimensionless
vibrational-bath interaction strength scaling factor. This implies
that all intramolecular vibrational modes will have the same capacity
for relaxation.

Equations of the model dynamics are obtained by applying the time-dependent
Dirac-Frenkel variational method \citep{Frenkel1931}. The main idea
behind Dirac-Frenkel variational method is that a parametrized trial
wavefunction $|\Psi\left(\boldsymbol{\xi}\left(t\right)\right)\rangle$
is varied so that the model Lagrangian $\mathcal{L}\left(t\right)$
is maintained at maxima (or minima). For this purpose, the time evolution
of every free parameter $\xi_{i}\left(t\right)$ is deduced using
the Euler-Lagrange equation
\begin{equation}
\frac{\text{d}}{\text{d}t}\left(\frac{\partial\mathcal{L}\left(t\right)}{\partial\dot{\xi_{i}}^{\star}\left(t\right)}\right)-\frac{\partial\mathcal{L}\left(t\right)}{\partial\xi_{i}^{\star}\left(t\right)}=0\ ,
\end{equation}
where $\dot{\xi_{i}}$ is the time derivative of $\xi_{i}$ and Lagrangian
$\mathcal{L}\left(t\right)$ of the model is given by 
\begin{align}
\mathcal{L}\left(t\right) & =\frac{\text{i}}{2}\left(\langle\Psi\left(\boldsymbol{\xi}\left(t\right)\right)|\dot{\Psi}\left(\boldsymbol{\xi}\left(t\right)\right)\rangle-\langle\dot{\Psi}\left(\boldsymbol{\xi}\left(t\right)\right)|\Psi\left(\boldsymbol{\xi}\left(t\right)\right)\rangle\right)\nonumber \\
 & -\langle\Psi\left(\boldsymbol{\xi}\left(t\right)\right)|\hat{H}|\Psi\left(\boldsymbol{\xi}\left(t\right)\right)\rangle\ .
\end{align}
The procedure results in a system of time-dependent equations for
parameters $\boldsymbol{\xi}\left(t\right)$, which minimize the deviation
of $|\Psi\left(\boldsymbol{\xi}\left(t\right)\right)\rangle$ from
the solution of the corresponding Schrödinger equation.

For this work, we define a Davydov $\text{D}_{2}$ \textit{ansatz}
superposition ($\text{sD}_{2}$) wavefunction
\begin{align}
|\Psi_{\text{sD}_{2}}\left(t\right)\rangle= & \sum_{n}^{N}\int_{\boldsymbol{x}^{\text{min}}}^{\boldsymbol{x}^{\text{max}}}\text{d}\boldsymbol{x}\varPhi_{n}\left(\boldsymbol{x},t\right)|n,\boldsymbol{x}\rangle\nonumber \\
 & \times\sum_{\alpha}^{M}\theta_{\alpha}\left(t\right)\prod_{p}^{P}|\lambda_{\alpha p}\left(t\right)\rangle\ .\label{eq:sd2}
\end{align}
The first product term of $\text{sD}_{2}$ defines all possible vibronic
states of the system with complex amplitudes $\varPhi_{n}\left(\boldsymbol{x},t\right)$.
The sum over index $n$ represents a superposition of electronic states,
while Q-dimensional integral represents coordinate basis states of
internal vibrational modes. Representation of the internal vibronic
states spans the whole space of the system states, and it can be used
to calculate dynamics of the system to an arbitrary precision. The
second product term defines possible states of the bath QHO modes.
Each mode $p$ of the bath is represented by a superposition of $M$
coherent states $|\lambda_{\alpha p}\left(t\right)\rangle\left(\alpha=1,2,\ldots,M\right)$
with CS displacements $\lambda_{\alpha p}\left(t\right)$, while each
superposition term $\alpha$ is parameterized by a complex amplitude
$\theta_{\alpha}\left(t\right)$. In general, a single CS is an eigenstate
of QHO annihilation operator $\hat{a}|\lambda\left(t\right)\rangle=\lambda\left(t\right)|\lambda\left(t\right)\rangle$,
whose displacement $\lambda\left(t\right)$ uniquely defines properties
of the oscillator wavepacket \citep{Rodney2000}. Interpretation of
CS displacement is especially straightforward in the coordinate and
momentum $\left(\chi,\rho\right)$ phase space with respective operator
expectation values being equal to
\begin{align}
\overline{\chi}\left(t\right) & =\sqrt{2}\text{Re}\lambda\left(t\right)\ ,\label{eq:exp_X}\\
\overline{\rho}\left(t\right) & =\sqrt{2}\text{Im}\lambda\left(t\right)\ .\label{eq:exp_P}
\end{align}
Per definition, the single CS wavepacket always remains Gaussian and
is centered in $\left(\overline{\chi}\left(t\right),\overline{\rho}\left(t\right)\right)$
phase space point at time $t$, thus, it follows trajectory defined
solely by the $\lambda\left(t\right)$. By considering superposition
of CSs, we allow for the wavepacket of each mode $p$ to be composed
of $M$ interfering Gaussian wavepackets. As such, the superposition
can represent non-Gaussian wavefunctions of the excited QHO states.
For the superposition length of $M=1$, representation of the bath
vibrational mode states by the $\text{sD}_{2}$ wavefunction is reduced
to the standard $\text{D}_{2}$ \textit{ansatz - }single CS wavepacket
representation. The $\text{sD}_{2}$ parameter $M$ allows to incrementally
increase accuracy of the bath modeling. Alternatively, combining amplitudes
of the system and the bath into a single amplitude, $\varPhi_{n}\left(\boldsymbol{x},t\right)\times\theta_{\alpha}\left(t\right)\rightarrow\Upsilon_{n\alpha}\left(\boldsymbol{x},t\right)$
would remove BOA, giving the most general and, presumably, the most
accurate \textit{ansatz} at a cost of significantly increased computational
effort \citep{Zhou2015a,Zhou2016,Wang2016}.

With the superposition given by $\text{sD}_{2}$ wavefunction (\ref{eq:sd2}),
normalization of the wavefunction $\langle\Psi_{\text{sD}_{2}}\left(t\right)|\Psi_{\text{sD}_{2}}\left(t\right)\rangle=1$
imposes
\begin{align}
\sum_{n}\int\text{d}\boldsymbol{x}\varPhi_{n}^{\star}\left(\boldsymbol{x},t\right)\varPhi_{n}\left(\boldsymbol{x},t\right) & =1\ ,\\
\sum_{\alpha,\beta}\theta_{\alpha}^{\star}\left(t\right)\theta_{\beta}\left(t\right)S_{\alpha\beta}\left(t\right) & =1\ ,
\end{align}
conditions, where $S_{\alpha\beta}\left(t\right)=\prod_{k}\langle\lambda_{\alpha k}\left(t\right)|\lambda_{\beta k}\left(t\right)\rangle$
is an overlap of $\alpha$ and $\beta$ CS product superposition terms.

Applying the Dirac-Frenkel variational method to $\text{sD}_{2}$
wavefunction (\ref{eq:sd2}) with the Hamiltonian operator (\ref{eq:ham-oper}),
we derived model equations of motion in a form of a system of implicit
differential equations, see Supplementary Information for the details.

\section{Results and Discussion\label{sec:Results_Disc}}

In this section, the approach described above is used to investigate
excitation energy relaxation dynamics between two excited electronic
states $|1\rangle$, $|2\rangle$ attached to PES in an avoided crossing
configuration. The electronic ground state $|0\rangle$ is included
solely to account for the system before an external excitation. The
electronic states are attached to a single internal vibrational mode
$Q=1$ (therefore, we drop index $q$) with Morse PES $V_{11}\left(x\right)$
for $|1\rangle$ and harmonic PESs $V_{00}\left(x\right)$ and $V_{22}\left(x\right)$
for $|0\rangle$ and $|2\rangle$ states, respecitvely
\begin{align}
V_{00}\left(x\right) & =\frac{\omega}{2}x^{2}\ ,\\
V_{11}\left(x\right) & =D_{0}\left(1-\text{e}^{-\frac{\left(x-d_{1}\right)}{\sqrt{2D_{0}}}}\right)^{2}\ ,\\
V_{22}\left(x\right) & =\frac{\omega}{2}\left(x-d_{2}\right)^{2}\ ,
\end{align}
with dissociation energy $D_{0}$ and frequency $\omega$. The PES
equilibrium points are displaced by $d_{1}=-1.35$ and $d_{2}=1.35$.
Excited state PESs are coupled by a linear non-adiabatic coupling
$V_{12}\left(x\right)=\frac{\omega}{10}x$, often refered to as a
\textit{vibronic} coupling. Throughout the paper we will use dimensionless
energy units by normalizing energies to $\omega$. In this scale,
we set electronic state energies to $\varepsilon_{1}=0$, $\varepsilon_{2}=5$,
PES $V_{00}\left(x\right)$ and $V_{22}\left(x\right)$ frequencies
to $\omega=1$, and dissociation energy to $D_{0}=40$. In the limit
of $D_{0}\rightarrow\infty$, $V_{11}\left(x\right)$ approaches harmonic
PES shape with frequency $\omega=1$. The coordinate $x$ space was
discretized in the interval from $x^{\text{min}}=-10$ to $x^{\text{max}}=12$
with equidistant step size of $\delta x=0.25$. The selected width
of $x$ space is large enough to include all non-negligible electron-vibrational
wavepacket amplitudes during its time evolution.

\begin{figure}
\includegraphics[width=8.25cm]{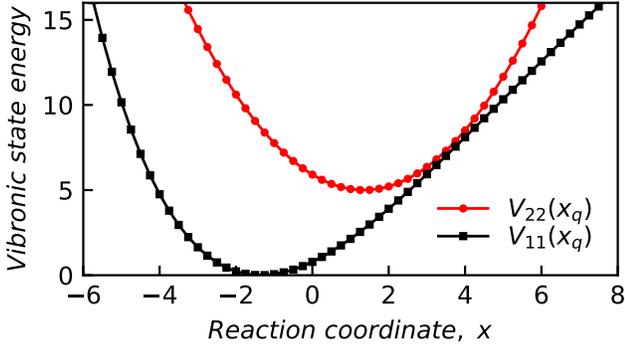}

\caption{Avoided crossing configuration of the first and second excited electronic
states $|1\rangle$, $|2\rangle$ attached to Morse $V_{11}\left(x\right)$
and harmonic $V_{22}\left(x\right)$ potential energy surfaces, respectively.
Ground state potential surface is not shown, while it is centered
at zero. Circle and square markers indicate considered coordinate
states $|x\rangle$ of the intramolecular vibrations. Optical excitation
from the ground state $|0\rangle$ to electronic state $|2\rangle$
results in Gaussian vibrational wavepacket centered at $x=0$ on $V_{22}\left(x\right)$
with variance $\sigma_{x}^{2}=1$. \label{fig:Potential-energy-surfaces}}
\end{figure}

We base these parameters on the typical energy scales found in organic
compounds present in Nature. Setting internal mode frequency to an
approximate frequency of carbon $\text{C=C}$ bond vibration $\omega=1500\ \text{cm}^{-1}$,
then the $|2\rangle\rightarrow|1\rangle$ internal conversion transition
energy gap is $\Delta\varepsilon_{21}=\varepsilon_{2}-\varepsilon_{1}=5\omega$,
which corresponds to an optical gap. Internal conversion energy gap
varies widely between molecular pigments, \textit{e. g.}, chlorophyll-A
$\text{Q}_{x}\text{-}\text{Q}_{y}$ energy gap is $\approx1.3\omega$
\citep{Shipman1976}, while $\text{S}_{2}\text{-}\text{S}_{1}$ energy
gap in carotenoids range from $\approx1\omega$ to $\approx5\omega$
depending on the carotenoid length \citep{Kosumi2009}. Rest of the
model parameters are kept quite arbitrary, since concrete parametrization
of both the chlorophyll and the carotenoid PESs are lacking. Note
that the ground electronic state does not couple to excited state
manifold via vibronic coupling and, thus, will be left out of the
analysis. The resulting avoided crossing PES configuration is shown
in Fig. (\ref{fig:Potential-energy-surfaces}).

Statistical properties of the bath fluctuations are represented by
the Ohmic spectral density function
\begin{align}
C_{n}''\left(w\right) & =\frac{w^{s}}{w_{\mathrm{c}}^{s-1}}\exp\left(-w/w_{c}\right),
\end{align}
with parameter $s=3$, cutoff frequency $w_{\mathrm{c}}=0.1$ and
the bath reorganization energy $\Lambda_{n}^{\text{bath}}=0.2$ for
each $n$. The frequency range of the bath vibrational modes $w\in\left[0.05,\ 2\right]$
was uniformly covered by $40$ modes with discretization step size
of $\text{d}w=0.05$. This setup is sufficiently dense to produce
the convergent dynamics and the interval of frequencies is wide enough
to cover all relevant resonances of the system-bath interactions.

The system and the bath interact via electronic-bath coupling (Eq.
\ref{eq:E-B cpl}) and one single vibrational-bath coupling term $k_{nqp}^{\left(1,2\right)}\hat{x}_{q}\hat{\chi}_{p}^{2}$
(see Eq. \ref{eq:vibr-bath_cpl}). \emph{I}\textit{.e.}, for simplicity
we set scaling factors to $\gamma^{\left(1,2\right)}=1$ and $\gamma^{\left(1,1\right)}=\gamma^{\left(2,1\right)}=0$.
Condition $\gamma^{\left(1,1\right)}=0$ guarantees that the bath
vibrational modes are retained as \emph{the normal modes, }while $\gamma^{\left(2,1\right)}=0$
implies that the double vibrational quanta absorption by the system
is not included.

\begin{figure}
\includegraphics[width=8.25cm]{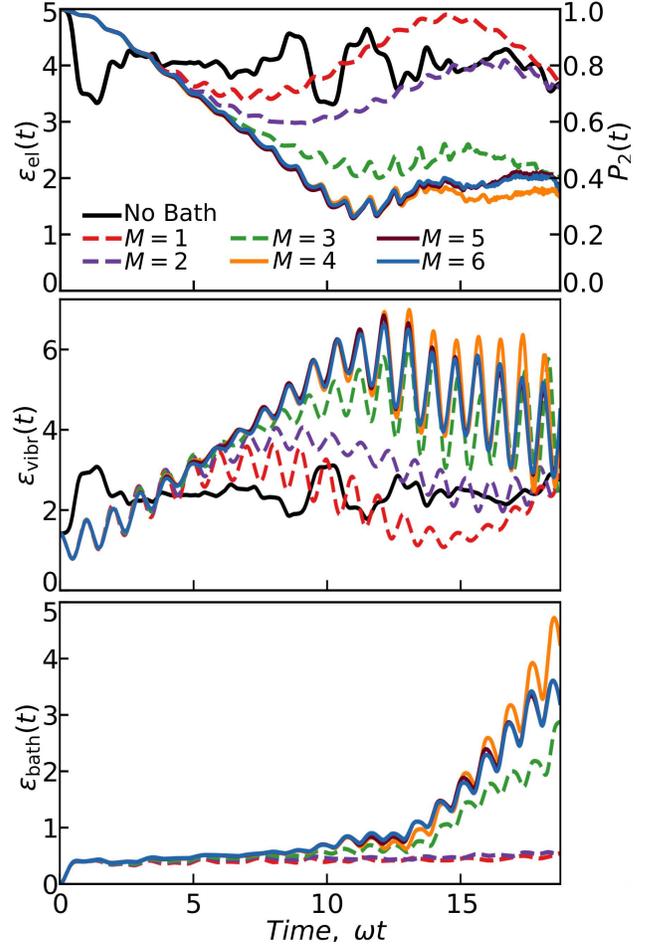}

\caption{Time dependence of ($a$) system electronic energy $\varepsilon_{\text{el}}$
and electronic state $|2\rangle$ population $P_{2}$, ($b$) system
vibrational energy $\varepsilon_{\text{vibr}}$ and ($c$) bath vibrational
energy $\varepsilon_{\text{bath}}$ calculated with no bath and superposition
length $M=1,\ldots,6$. \label{fig:td-energies}}
\end{figure}

Initial condition of the system and the bath are taken to correspond
to the lowest energy states. Assuming that either transition $|0\rangle\rightarrow|1\rangle$
is optically forbidden or is off-resonant, the optical excitation
by an external field is modeled using the Franck-Condon ground $|0\rangle$
to excited $|2\rangle$ state electronic transition. This corresponds
to the projection of the system ground state wavepacket into the 2-nd
excited state potential surface, setting $\varPhi_{2}\left(x,0\right)=\frac{1}{\sqrt{2\pi}}e^{-\frac{x^{2}}{2}}$
and $\varPhi_{1}\left(x,0\right)=0$. Wavepackets of the bath are
Gaussian as well, and they can be exactly represented by a single
CS. Correspondingly, we choose to set initial amplitudes to $\theta_{1}\left(0\right)=1$,
$\theta_{2\ldots M}\left(0\right)=0$ and CS displacements to $\lambda_{\alpha p}\left(0\right)=0$
for every combination of $\alpha,p$ indices: at $t=0$ there are
$M$ degenerate CSs, while only the first $\alpha=1$ CS amplitude
is non-zero, and all QHO wavepackets are centered in their respective
coordinate-momentum phase space $\left(\chi_{p}=0,\rho_{p}=0\right)$.
Notice, that when $M>1$, the same bath initial condition can be achieved
by parametrizing coherent states differently, we will look at it later
in this work.

\subsection{Following system energy and dissipation\label{subsec:System-excitation-energy}}

To track excitation energy relaxation within the system itself and
energy exchange between the system and the bath, we look at dynamics
of system electronic, vibrational energies $\varepsilon_{\text{el}}\left(t\right)$,
$\varepsilon_{\text{vibr}}\left(t\right)$ and bath vibrational energy
$\varepsilon_{\text{bath}}\left(t\right)$ defined as
\begin{align}
\varepsilon_{\text{el}}\left(t\right)= & \sum_{n}\varepsilon_{n}P_{n}\left(t\right)\ ,\\
\varepsilon_{\text{vibr}}\left(t\right)= & \sum_{n}\int\text{d}x\varPhi_{n}^{\star}\left(x,t\right)\left(V_{nn}\left(x\right)-\frac{\omega}{2}\frac{\partial^{2}}{\partial x^{2}}\right)\varPhi_{n}\left(x,t\right)\nonumber \\
 & +\sum_{n,m}^{n\neq m}\int\text{d}x\varPhi_{n}^{\star}\left(x,t\right)V_{nm}\left(x\right)\varPhi_{m}\left(x,t\right)\label{eq:E_vibr}\\
\varepsilon_{\text{bath}}\left(t\right)= & \sum_{\alpha,\beta,p}\theta_{\alpha}^{\star}\left(t\right)\theta_{\beta}\left(t\right)\lambda_{\alpha p}^{\star}\left(t\right)\lambda_{\beta p}\left(t\right)S_{\alpha\beta}\left(t\right)\ ,
\end{align}
with $P_{n}\left(t\right)=\int\text{d}x\left|\varPhi_{n}\left(x,t\right)\right|^{2}$
being the $n$-th electronic state population. For consistency with
the system Hamiltonian (Eq. \ref{eq:H_sys}), we include non-adiabatic
coupling $V_{nm}\left(x\right)$ energy in the definition of $\varepsilon_{\text{vibr}}\left(t\right)$,
also, for simplicity, we exclude QHO zero-point energy from the bath
energy $\varepsilon_{\text{bath}}\left(t\right)$.

In Fig. (\ref{fig:td-energies}) we present time dependence of the
system, bath energies and initially occupied electronic state $|2\rangle$
population $P_{2}$ calculated with superposition length $M=1,\ldots,6$.
For reference we also plot system energy dynamics of an isolated system.
Notice that, because excitation energy of the state $|1\rangle$ is
$\varepsilon_{1}=0$, the total electronic energy is a function of
just $|2\rangle$ electronic state population, $\varepsilon_{\text{el}}\left(t\right)=\varepsilon_{2}P_{2}\left(t\right)$.

In the case of an isolated system, non-trivial oscillations between
the system electronic and vibrational energy are observed (internal
conversion due to non-adiabiatic PES coupling $V_{12}\left(x\right)$),
however, only about 25$\%$ of the electronic state $|2\rangle$ population
$P_{2}$ transfers to $|1\rangle$ state and a large amount of the
transfered population from the state $|1\rangle$ is then repeatedly
transfered back to $|2\rangle$ state -- internal conversion is reversible.
Now, let us also include the bath and couple it to the system. In
the case of the bath wavefunction representation by $M=1$ superposition
terms, character of the system energy oscillations changes: it now
displays harmonic, reversible behavior with a period of $\tau_{\text{IC}}\approx15\omega t$.
Additionally, $\varepsilon_{\text{el}}$ and $\varepsilon_{\text{vibr}}$
are also modulated with a period of $\tau_{\text{IV}}\approx\omega t$,
yet, with a smaller modulation amplitude. Also, no appreciable vibrational
energy exchange between the system and bath modes is observed, the
slight increase in the bath energy is solely due to the electron-bath
coupling induced bath reorganization. By increasing superposition
length to $M=2$, system electronic and vibrational energies no longer
simpy oscillate, but some of the electronic energy is irreversibly
converted into the system vibrational energy. Still, no significant
energy dissipation to the bath occurs. Taking $M=3$, non-negligible
energy exchange between the system and the bath vibrational modes
begins. Considering even more superposition terms, non-reversible
internal conversion and dissipation effects become further pronounced
and converge at $M=5$. The convergent non-reversible internal conversion
occurs on a time scale of $\tau_{\text{IC}}$ with 60$\%$ of the
initially occupied $|2\rangle$ state population relaxed to the $|1\rangle$
state, which is followed by the system vibrational energy dissipation
to the bath.

The drastic change in the behavior of the $\varepsilon_{\text{el}}$
and $\varepsilon_{\text{vibr}}$ energies, when the system becomes
coupled to the bath, is induced by the bath vibrational mode action
on the evolution of the system electron-vibrational wavepacket due
to the vibrational-bath coupling term $\propto k_{nqp}^{\left(1,2\right)}\hat{x}_{q}\hat{\chi}_{p}^{2}$.
For an isolated system, internal conversion dynamics are decided solely
by the free evolution and mixing of the electron-vibration wavepackets
on $V_{11}\left(x\right)$ and $V_{22}\left(x\right)$ PES. By coupling
the system to the bath, electron-vibrational wavepacket evolution
becomes influed by the motion of the bath vibrational modes. By looking
at the visualization of the electron-vibrational wavepacket evolution,
presented in Supplementary Information, we found that vibrational-bath
coupling effectivelly reduces oscillation amplitude of the electron-vibrational
wavepacket on $V_{22}\left(x\right)$, making it harder to reach the
avoided crossing area ($x\approx4$) between the $V_{11}\left(x\right)$
and $V_{22}\left(x\right)$, however, on each oscillation of the $V_{22}\left(x\right)$
PES electron-vibrational wavepacket with a period of $\tau_{\text{IV}}$,
a small amount of wavepacket is still transfered to $V_{11}\left(x\right)$.
In the case of $M=1$, for the first $\frac{1}{2}\tau_{\text{IC}}$
after excitation, we observe a gradual population transfer from the
$|2\rangle$ state to the $|1\rangle$ state with the reversed process
occuring for the following $\frac{1}{2}\tau_{\text{IC}}$. In the
convergent case of $M=5$, for roughly the full period of $\tau_{\text{IC}}$
we observe analogous population transfer from the $|2\rangle$ state
to $|1\rangle$, however, now the generated system vibrational energy
is non-reversibly dissipated to the bath, instead of being converted
back into the electronic $|2\rangle$ state energy.

The total lack of vibrational energy exchange between the system and
the bath vibrational modes at $M=1$ suggests that the simple $\text{D}_{2}$
\textit{ansatz} is incapable of representing any QHO states necessary
to absorb vibrational energy due to quadratic vibrational-bath coupling
term $\propto k_{nqp}^{\left(1,2\right)}\hat{x}_{q}\hat{\chi}_{p}^{2}$.
Meanwhile, superposition of CSs allows for the formation of QHO non-zero
vibrational energy state wavepackets and to absorb vibrational energy
from the system.

\begin{figure}
\includegraphics[width=8.25cm]{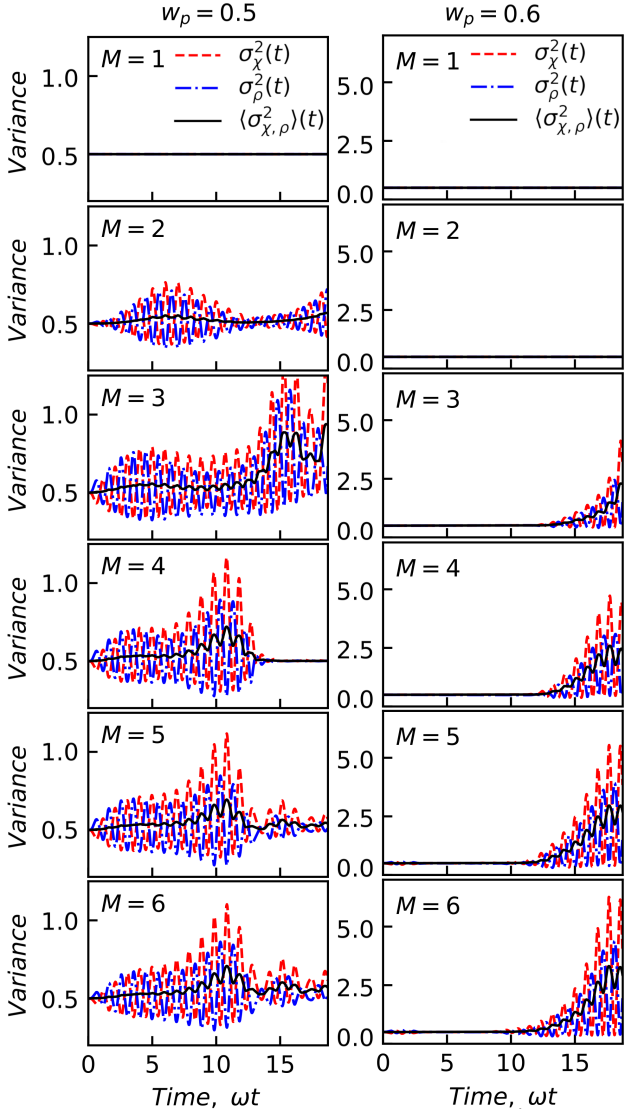}

\caption{Time dependence of frequency $\omega_{p}=0.5$ (left column) and $\omega_{p}=0.6$
(right column) bath vibrational mode coordinate $\sigma_{\chi}^{2}\left(t\right)$,
momentum $\sigma_{\rho}^{2}\left(t\right)$ variances and their arithmetic
average $\sigma_{\chi,\rho}^{2}\left(t\right)$ calculated with superposition
length $M=1,\ldots,6$. \label{fig:td-std}}
\end{figure}

To evaluate characteristics of the bath wavepackets, we have computed
coordinate, momentum variances and their arithmetic average for a
selected set of bath vibrational modes
\begin{align}
\sigma_{\chi_{p}}^{2}\left(t\right) & =\overline{\chi_{p}^{2}\left(t\right)}-\overline{\chi_{p}\left(t\right)}^{2}\ ,\label{eq:var_x}\\
\sigma_{\rho_{p}}^{2}\left(t\right) & =\overline{\rho_{p}^{2}\left(t\right)}-\overline{\rho_{p}\left(t\right)}^{2}\ ,\label{eq:var_p}\\
\left\langle \sigma_{\chi_{p},\rho_{p}}^{2}\right\rangle \left(t\right) & =\frac{1}{2}\left(\sigma_{\chi_{p}}^{2}\left(t\right)+\sigma_{\rho_{p}}^{2}\left(t\right)\right)\ ,\label{eq:var_mean}
\end{align}
where $\overline{\mathcal{O}_{p}\left(t\right)}=\langle\Psi_{\text{sD}_{2}}\left(\boldsymbol{x},t\right)|\hat{\mathcal{O}}_{p}|\Psi_{\text{sD}_{2}}\left(\boldsymbol{x},t\right)\rangle$
is an expectation value of operator $\hat{\mathcal{O}}_{p}$. We have
chosen to look at two modes with frequencies close to half of the
electronic energy gap, $w_{p}\approx\frac{\omega}{2}$, as it is the
frequency of the expected resonance band created by the quadratic
vibrational-bath coupling. Time dependence of frequency $w_{p}=0.5$
and $w_{p}=0.6$ bath vibrational modes variances calculated with
$M=1,\ldots,6$ are shown in Fig. (\ref{fig:td-std}).

In the case of $M=1$, both coordinate and momentum variances are
equal to $0.5$ and, as expected, they do not change in time, because
the wavepacket of each mode remains strictly Gaussian. Taking $M=2$,
the coordinate and momentum variances of the mode with $w_{p}=0.5$
oscillate almost harmonically, indicating that the wavepacket remains
almost Gaussian, but it is successively squeezed along $\chi_{p}$
and $\rho_{p}$ axes (behavior characteristic of the squeezed coherent
states); no significant variance change for $w_{p}=0.6$ is observed.
Considering $M=3$, variance oscillations of the mode with $w_{p}=0.5$
are no longer harmonic, \textit{i.e.}, oscillation amplitude maximum
of $\sigma_{\chi_{p}}^{2}$ exceeds that of $\sigma_{\rho_{p}}^{2}$,
implying, that Gaussian wavepacket is asymmetrically squeezed; mode
$w_{p}=0.6$ variances oscillate are now also observed. Including
more superposition terms, pattern of the variance oscillations continue
to change, and, in accordance with energy dynamics, superposition
of $M=5$ provide convergent dynamics, with both modes displaying
anharmonic variance oscillations.

Variance oscillation amplitudes of $w_{p}=0.6$ mode is about $5$
times greater than that of $w_{p}=0.5$ mode, suggesting, that the
former mode must lie in an effective resonance band for considered
vibrational-bath coupling and it is responsible for absorbing the
major part of vibrational energy dissipated from the system to the
bath. The latter mode is off-resonant and contributes less to the
vibrational energy absorption. Also, variance oscillation pattern
of $w_{p}=0.6$ mode closely resembles that of $\varepsilon_{\text{bath}}\left(t\right)$
in Fig. (\ref{fig:td-energies}), further providing evidence, that
this mode is the main absorber of the system vibrational energy.

Additionally, the variance averages $\left\langle \sigma_{\chi,\rho}^{2}\right\rangle $
of both modes are not static and exceed variance average of the initially
prepared Gaussian wavepacket, implying, that wavepackets broaden.
This is in accordance to uncoupled QHO variance analytical solution,
which states that $\left\langle \sigma_{\chi,\rho}^{2}\right\rangle =\frac{1+2k}{2}$
is linearly proportional to QHO eigenstate occupation number $k$.
In our case, bath QHO high occupation number states are accessed by
absorbing vibrational energy from the system.

\subsection{Lifting coherent state degeneracy}

In a previous section the initial bath state, corresponding to the
lowest energy QHO states, was represented by degenerate CSs, \textit{i.e.},
all vibrational mode $p$ CS displacements were the same $\lambda_{\alpha p}\left(t=0\right)=0$.
However, because we set only the first superposition term amplitude
to be non-zero $\theta_{1}\left(0\right)=1$, $\theta_{2\ldots M}\left(0\right)=0$
we can choose any initial displacement $\lambda_{\alpha\geq2,p}\left(0\right)$
value, without actually changing the bath initial condition. Therefore,
now we investigate the effects of setting $\lambda_{\alpha\geq2,p}\left(0\right)\neq\lambda_{\alpha=1,p}\left(0\right)$,
which will lift initial CS degeneracy, on the model dynamics and convergence.

\begin{figure}
\includegraphics[width=6cm]{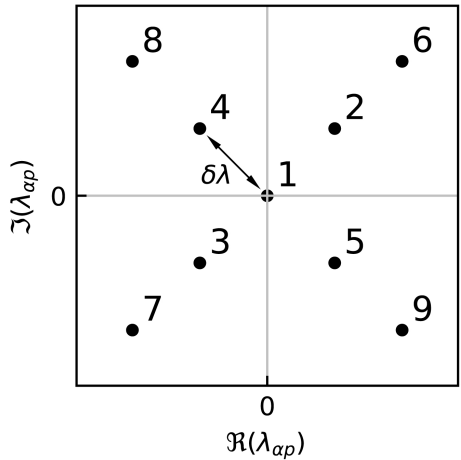}

\caption{Scheme of coherent state displacement $\lambda_{\alpha p}\left(t=0\right)$
arrangement of $\alpha=1,\ldots,9$ states for all $p$. Free parameter
$\delta\lambda$ determines separation between the nearest Gaussian
wavepacket states $\alpha$.\label{fig:init-scheme}}
\end{figure}

With our choice of $\theta_{\alpha}\left(0\right)$, displacement
$\lambda_{\alpha=1,p}\left(0\right)$ define initial $p$-th vibrational
mode Gaussian wavepacket position in the coordinate-momentum phase
space $\left(\chi_{p},\rho_{p}\right)$, see Eqs. (\ref{eq:exp_X}),
(\ref{eq:exp_P}), while displacements $\lambda_{\alpha\geq2,p}\left(0\right)$
define additional Gaussian wavepacket states of $p$-th vibrational
mode in phase space, though, they carry zero amplitudes initially.
Ideally, we would like to cover as much of phase space as possible
with additional states, yet, keep them close enough to each other
for their wavepackets to overlap, and centered around the initially
populated state $\alpha=1$. Therefore, we chose to arrange initial
CS displacements $\lambda_{\alpha p}\left(0\right)$ in a cross-like
pattern, see Fig. (\ref{fig:init-scheme}), while keeping $\lambda_{\alpha=1,p}\left(0\right)=0$
centered in phase space. Displacement pattern is reproduced by an
expression 
\begin{equation}
\lambda_{\alpha p}\left(0\right)=\frac{\delta\lambda}{\sqrt{2}}\left(1+\left\lfloor \frac{\alpha-2}{4}\right\rfloor \right)\left(\left(-1\right)^{\left\lfloor \frac{\alpha}{2}\right\rfloor +\alpha+1}+\text{i}\left(-1\right)^{\alpha}\right)\ ,
\end{equation}
where $\left\lfloor \mathcal{O}\right\rfloor $ is a floor function
of $\mathcal{O}$, and the parameter $\delta\lambda$ determines the
separation between the nearest Gaussian wavepacket states $\alpha$,
allowing to control their overlap. Separation of $\delta\lambda=0$
reproduces initial bath state basis used in Section. \ref{subsec:System-excitation-energy}.

\begin{figure}
\includegraphics[width=8.25cm]{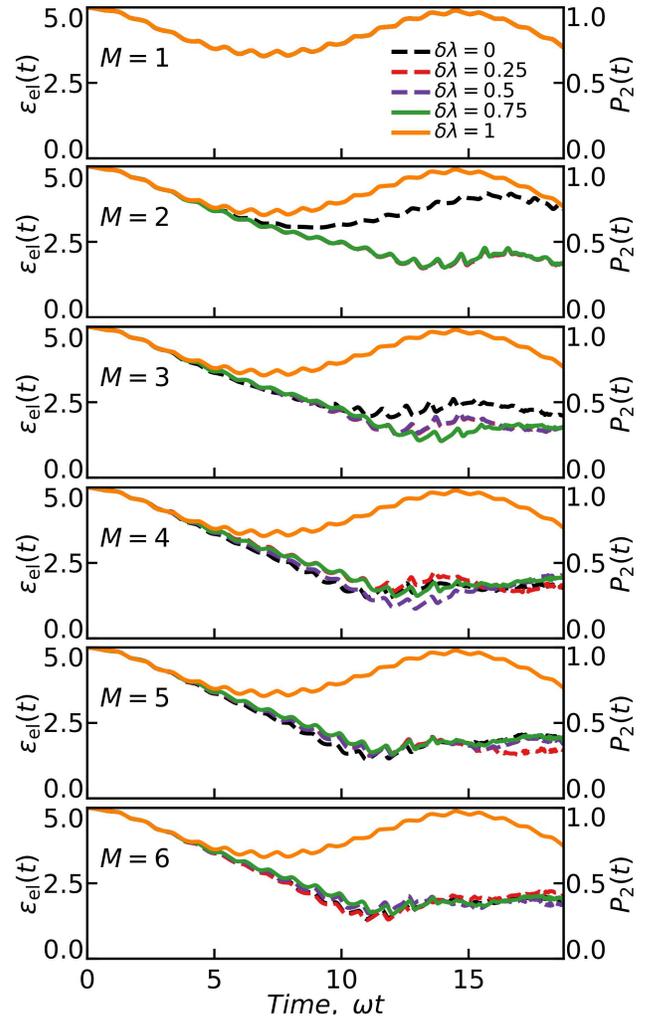}

\caption{Time dependence of the system electronic energy $\varepsilon_{\text{el}}\left(t\right)$
calculated with superposition length $M=1\ldots6$ and separation
$\delta\lambda=0\ldots1$ between the nearest coherent states. \label{fig:td-init}}
\end{figure}

In Fig. (\ref{fig:td-init}) we display the time dependence of the
system electronic energy $\varepsilon_{\text{el}}\left(t\right)$
calculated with superposition length $M=1,\ldots,6$ and separation
$\delta\lambda=0,\ldots,1$. In the case of $M=2$, state separations
in the range of $\delta\lambda=0.25,\ldots,0.75$ provide identical
and already semi-convergent result, as compared to the $M=5$ case,
while the degenerate state $\delta\lambda=0$ case only slightly differs
from the $M=1$. Large separation of $\delta\lambda=1$ performs the
worst and do not differ from the $M=1$ case at all, suggesting, that
the CS wavepackets no longer sufficiently overlap to allow formation
of necessary QHO state wavepackets. By further increasing the number
of superposition terms $M$, the dynamics calculated with $\delta\lambda=0.25,\ldots,0.75$
separations provide similar results (at the same $M$), suggesting,
that a small initial state separation does not drastically change
long term dynamics. Eventually, by considering $M=6$ terms, dynamics
with no $\delta\lambda=0$ and small $\delta\lambda=0.25,\ldots,0.75$
separations provide identical convergent result. On the other hand,
if separation is too large $\delta\lambda=1$, dynamics do not converge
at all, independent of a number of $M$ terms considered.

\subsection{Discussion}

To model vibrational energy exchange between the system ($q$) and
the bath ($p$) vibrational modes, we have included vibrational-bath
coupling term $\hat{H}_{\text{V-B}}\propto\hat{x}_{q}\hat{\chi}_{p}^{2}$.
It is the simplest coupling term that still maintains vibrational
modes as normal modes, since the bilinear coupling term $\hat{H}_{\text{V-B}}\propto\hat{x}_{q}\hat{\chi}_{p}$
would only redistribute oscillation amplitudes among vibrational modes
and by performing a unitary transformation we would obtain \textit{uncoupled}
normal modes.

Effects of quadratic electronic-phonon coupling term $\hat{H}_{\text{E-B}}\propto\hat{\chi}_{p}^{2}$
on time-resolved fluorescence of a single absorber have been investigated
by Chorošajev, \textit{et al. }\citep{Chorosajev2017a}. They represented
bath QHO states by a single squeezed coherent state (SCS) and were
able to account for spectral signatures of absorption from the hot
ground state, and the breaking of the absorption and relaxed fluorescence
mirror symmetry, \textit{i.e.}, the effects lacking in CS representation.
SCS approach was also applied to model Morse vibrational modes and
was shown to lead to wavepacket reorganization due to PES anharmonicity
\citep{Abramavicius2018b}. The ability of a single SCS to represent
QHO wavepackets is greater than a single CS, as it allows to model
symmetric $\sigma_{\chi}^{2}$ and $\sigma_{\rho}^{2}$ variance oscillations,
yet, it is still limited to just Gaussian wavepackets. Our more general
approach revealed, that quadratic vibrational-bath coupling not only
induces \textit{assymetric} QHO variance oscillations, but also broadens
wavepacket in coordinate and momentum phase space. Correct representation
of both of these effect by a single SCS is inherently impossible.
We believe that these effects would also be present in models with
quadratic electronic-phonon coupling term. Interestingly, superposition
of $M=2$ terms produced rather symmetric $\sigma_{\chi}^{2}$ and
$\sigma_{\rho}^{2}$ variance oscillations and could perhaps be an
alternative to using SCS for other applications. Additionally, we
found that solely \textit{linear} electron-bath coupling model does
not induce bath vibrational mode wavepacket variance changes, thus,
the bath state representation by a single CS is sufficient (not shown).

Although the considered CSs are dynamical, \textit{i.e.}, CS displacements
$\lambda_{\alpha p}\left(0\right)$ evolve in time, choosing more
appropriate CSs could perhaps better accomodate QHO wavepacket at
early times, leading to faster convergence and less computational
effort. We found that the small and medium separations $\delta\lambda=0,\ldots,0.75$
between the nearest states provided semi-convergend dynamics at superposition
of just $M=2$ terms, however, $M=5$ terms were required to obtain
a fully convergent result, independent of the separation. Interestingly,
if initial separation is too large $\delta\lambda=1$, it stays too
large at all times, indicated by identical dynamics obtained with
$M=1,\ldots,6$ terms. As for computational effort, while keeping
the same $M$, zero separation case $\delta\lambda=0$ required the
least computational effort and increasing separation only slowed down
calculations (not shown).

Regarding the form of the wavefunction, $\text{multi-D}_{2}$ \textit{ansatz}
does not use BOA and represents both system and bath vibrational mode
states using CSs. By increasing superposition length of $\text{multi-D}_{2}$,
representation accuracy (and numerical effort) of both the system
and the bath vibrational states increases equally. Lipeng, \textit{et
al.} have simulated pyrazene electron-vibrational wavepacket relaxation
through conical intersection using the $\text{multi-D}_{2}$ \textit{ansatz}
\citep{Chen2019} by considering two-level system with 4 internal
vibrational modes, 20 bath modes were \textit{linearly} coupled to
electronic states \citep{Worth1998}. Dynamics, obtained by including
more than 40 $\text{multi-D}_{2}$ superposition terms, agreed well
with those obtained using the state-of-the-art MCTDH method. It is
well known that modeling of internal conversion requires non-BOA representation
of the entangled system electron-vibrational wavepacket, for which
$\text{multi-D}_{2}$ is well suited, however, question remains of
whether one can apply BOA to separate system and bath wavefunctions,
and have non-reversible internal conversion, and whether representation
of the separated bath wavefunction need to be more complex than just
a single CS.

The $\text{sD}_{2}$ \textit{ansatz} defined here is of BOA structure,
however, the most important DOFs for internal conversion, \textit{i.e.},
entangled system electronic states and internal vibrational modes,
are treated formally exactly. Using the $\text{sD}_{2}$ \textit{ansatz},
we found that it is capable of modeling non-reversible internal conversion
in an avoided crossing configuration and that internal conversion
induced dynamics of the system electron-vibrational wavepacket is
highly dependent on the complexity of the bath wavefunction representation.
The simplest approach of Davydov $D_{2}$ ansatz with CS ($M=1$),
or even SCS (simillar to $M=2$), is not sufficient, because of their
limited ability in repersenting complex QHO wavepackets. To obtain
convergent non-reversible internal conversion dynamics of a model
molecule with electronic state energy gap in an optical band, we had
to include superposition of at least $M=5$ CS terms. Non-reversibility
is induced by the system vibrational energy dissipation to the bath
vibrational modes. The full \textit{ab initio} model of pyrazene \citep{Sala2015}
suggests an alternative excitation relaxation pathway via conical
intersection between the optically dark $A_{u}\left(n\pi^{\star}\right)$
state and pyrazene ground state, theoretical description of which
requires treatement of quadratic and higher order vibronic coupling,
in the form of Eq. (\ref{eq:vibr-bath_cpl}). Therefore, results of
this work could be of interest.

We considered $\text{0\ K}$ temperature limit. Stochastic extensions
of the Dirac-Frenkel variational method have been developed to account
for the temperature of the bath when using both single \citep{Chorosajev2016b}
and multi \citep{Wang2017a} variants of Davydov $\text{D}_{2}$ \textit{ansatz}.
These extensions average initial CS displacement realizations by sampling
QHO canonical ensemble density matrix. This correctly accounts for
initial canonical ensemble statistics, however, QHO wavepackets of
each realization is represented by a Gaussian, irrespective of the
temperature. This is fine in the \textit{linear} system-bath coupling
regime, as only the coordinate averages of QHOs are of interest, meanwhile,
when considering higher order coupling terms, one would have to correctly
account for the initially non-Gaussian wavepacket for each realization.

\section{Conclusion\label{sec:Conclusion}}

In summary, the presented theory allows to investigate non-reversible
molecular internal conversion dynamics with simultaneous system thermal
energy dissipation to the bath. We defined $\text{sD}_{2}$ \textit{ansatz},
which represents the most essential states for internal conversion,
\textit{i.e.}, entangled electron-vibrational wavepacket states, formally
exactly, while bath quantum harmonic oscillator states were expanded
in a superposition of coherent states. To have thermal energy dissipation
to the bath, we included non-linear coupling term $\hat{H}_{\text{V-B}}\propto\hat{x}_{q}\hat{\chi}_{p}^{2}$
between the system and the bath vibrational modes. Using non-adiabatically
coupled three-site model, we showed that non-linear system-bath coupling
induced non-reversible internal conversion requires highly non-Gaussian
bath quantum harmonic oscillator wavepacket representation, as well
as, that non-linear coupling results in a broadened and asymmetrically
squeezed wavepacket. We argue that these effects are, per definition,
not possible to model with simple Davydov $\text{D}_{2}$ \textit{ansatz,}
while squeezed coherent state representation is insufficient. Also,
that coupling terms \textit{linearly} proportional to bath vibrational
mode coordinate $\hat{H}\propto\hat{\chi}_{p}$ does not induce wavepacket
changes, thus, bath state representation by a single coherent state
is sufficient. Additionally, we compared model dynamics and convergence
with degenerate and non-degenerate initial coherent states and found
that the degenerate case provided the same convergent result as the
non-degenerate situation, however, required less computational effort.
The presented approach is general and could be used to model effects
of other types of non-linear system-bath couplings.

\section*{Acknowledgement}

This research was funded by the European Social Fund under the No
09.3.3-LMT-K-712 \textquotedblleft Development of Competences of Scientists,
other Researchers and Students through Practical Research Activities\textquotedblright{}
measure. T. M. was supported by the Czech Science Foundation (GACR)
grant no. 17-22160S. Computations were performed on resources at the
High Performance Computing Center \quotedblbase HPC Sauletekis\textquotedblleft{}
in Vilnius University Faculty of Physics.

\section*{Conflicts of interest}

There are no conflicts of interest to declare.

\bibliographystyle{pnas}
\addcontentsline{toc}{section}{\refname}\bibliography{sD2}

\end{document}